\documentstyle[12pt,psfig]{article}
\begin{document}
\begin{center}
\Large{\bf Photon production in relativistic heavy-ion collisions
using rates with two-loop calculations from quark matter}
\vskip 0.2in

\large{Dinesh Kumar Srivastava}
\vskip 0.2in

\large{\em Variable Energy Cyclotron Centre,\\
 1/AF Bidhan Nagar, Calcutta 700 064, India}

\vskip 0.2in

\large{Erratum: Eur. Phy. J. C 10 (1999) 487.}

\vskip 0.2in

\end{center}

It has been shown recently~\cite{MT} shown that the values of $J_T$ and $J_L$
which appear in Eq.(2) and (5) of the above paper and which are
taken from the work of Aurenche et al\cite{pat} are too large by a factor of 4.
Correcting for this changes the Fig.~1--4 (see below).  The basic result
of the paper, that the emissions from the quark matter can outshine those
from the hadronic matter when the photon rates upto two-loop level are used,
remains valid though the range of $p_T$ over which it happens is reduced 
to $p_T <$ 0.5, 1, and 2 GeV/$c$ respecitively at SPS, RHIC, and LHC energies.

\newpage

\begin{figure}
\psfig{file=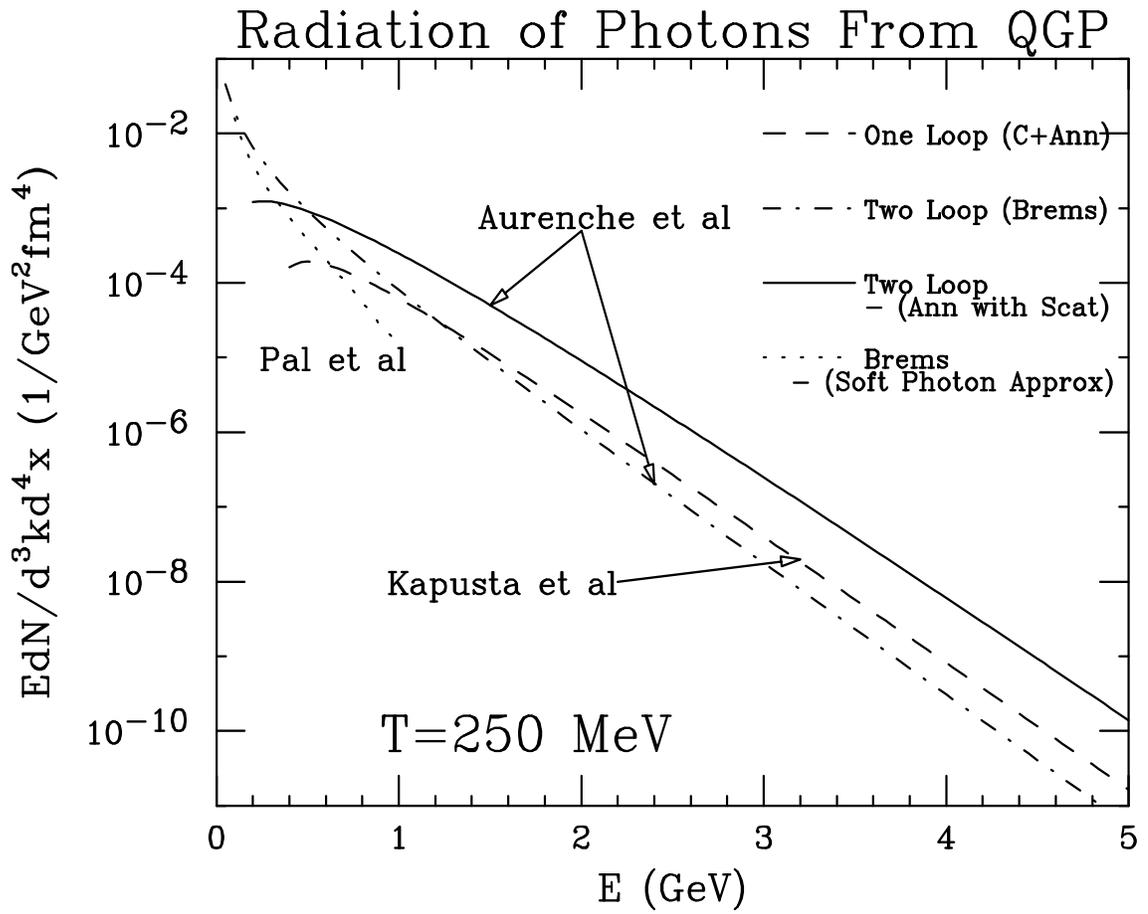,height=12cm,width=15cm}
\vskip 0.1in
\caption{ Radiation of photons from various processes in the quark
matter at $T=$ 250 MeV}
\end{figure}

\newpage

\begin{figure}
\psfig{file=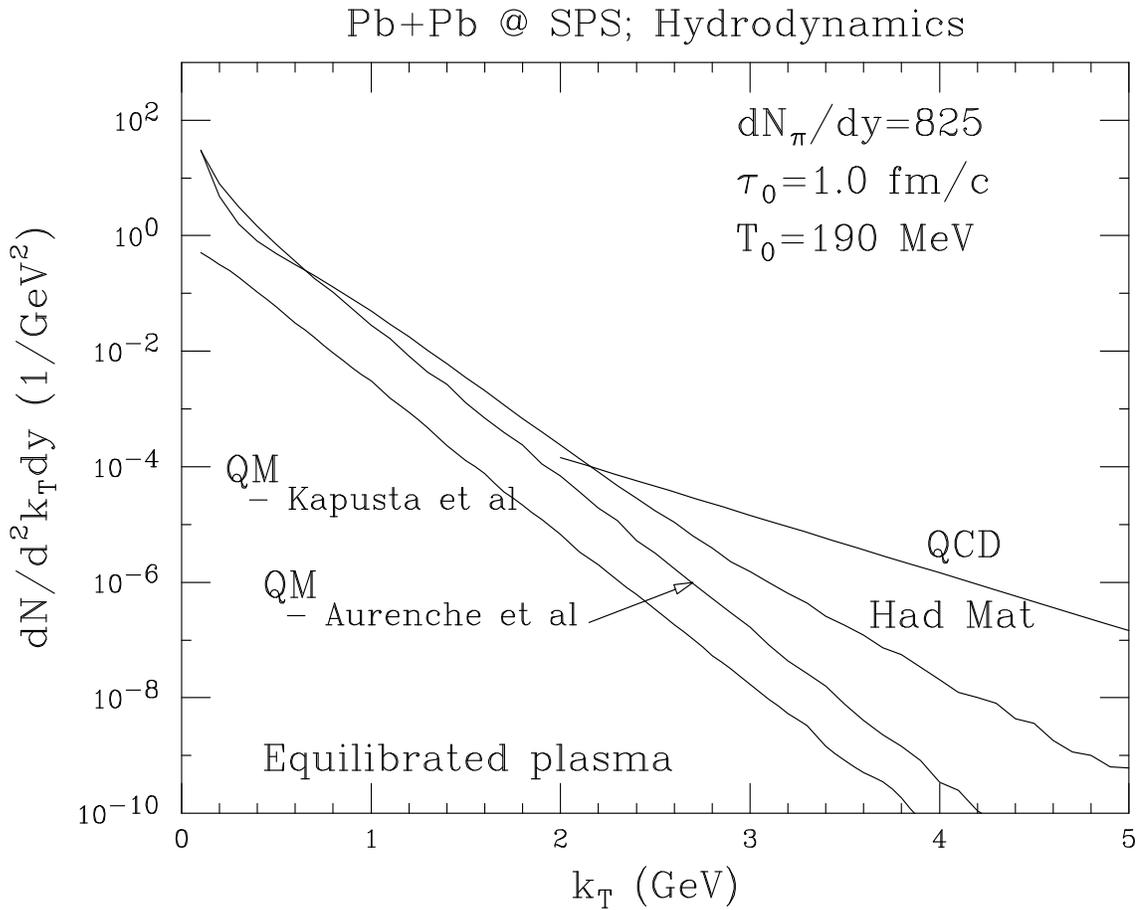,height=12cm,width=15cm}
\vskip 0.1in
\caption{ Radiation of photons from central collision of lead nuclei 
at SPS energies from the hadronic matter (in the mixed phase and the
hadronic phase) and the quark matter (in the QGP phase and the mixed
phase).
The contribution of the quark matter while using the
rates obtained by Kapusta et al and Aurenche et al,
and those from hard QCD processes 
 are shown separately
}
\end{figure}
\newpage
\begin{figure}
\psfig{file=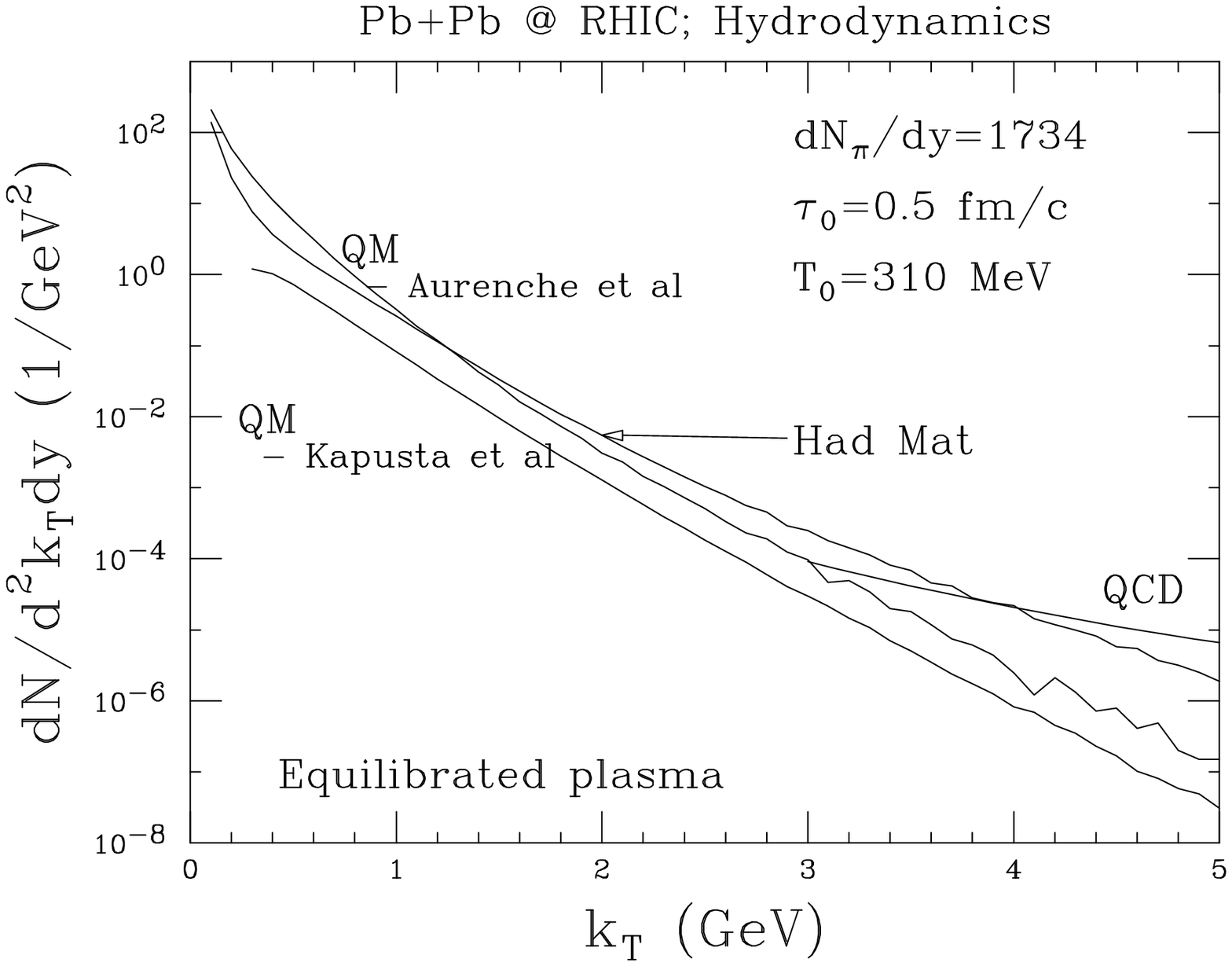,height=12cm,width=15cm}
\vskip 0.1in
\caption{ Same as Fig.~2 for RHIC energies.
}
\end{figure}
\newpage
\begin{figure}
\psfig{file=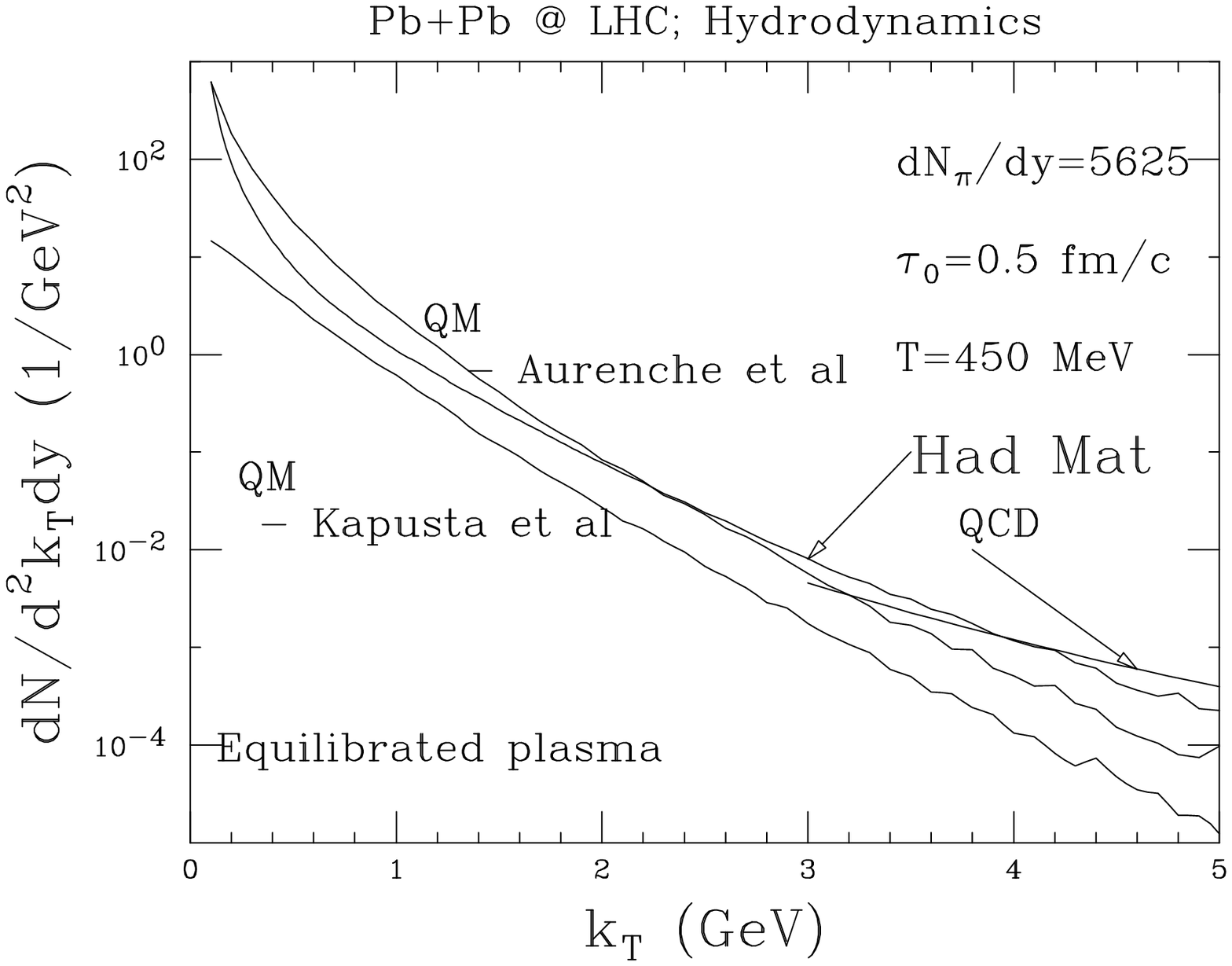,height=12cm,width=15cm}
\vskip 0.1in
\caption{ Same as Fig.~2 for LHC energies. 
}
\end{figure}
\end{document}